\def\year{2020}\relax
\documentclass[letterpaper]{article} 
\usepackage{aaai20}  
\usepackage{times}  
\usepackage{helvet} 
\usepackage{courier}  
\usepackage[hyphens]{url}  
\usepackage{graphicx} 
\usepackage{graphicx}  
\usepackage{amsfonts}
\usepackage{amsmath}
\usepackage{bm}
\usepackage{booktabs}
\usepackage{multirow}
\usepackage{subfigure}
\title{Rumor Detection on Social Media with Bi-Directional Graph Convolutional Networks}
\author{
Tian Bian,\textsuperscript{\rm 1,2} 
Xi Xiao,\textsuperscript{\rm 1} 
Tingyang Xu,\textsuperscript{\rm 2} 
Peilin Zhao,\textsuperscript{\rm 2}
Wenbing Huang,\textsuperscript{\rm 2}
Yu Rong,\textsuperscript{\rm 2} 
Junzhou Huang\textsuperscript{\rm 2}\\
\textsuperscript{\rm 1}Tsinghua University\\ 
\textsuperscript{\rm 2}Tencent AI Lab\\ 
bt18@mails.tsinghua.edu.cn,xiaox@sz.tsinghua.edu.cn, hwenbing@126.com, yu.rong@hotmail.com\\
\{tingyangxu, masonzhao, joehhuang\}@tencent.com \\
}
 
\begin{document}

\maketitle

\begin{abstract}
Social media has been developing rapidly in public due to its nature of spreading new information, which leads to rumors being circulated. 
Meanwhile, detecting rumors from such massive information in social media is becoming an arduous challenge. 
Therefore, some deep learning methods are applied to discover rumors through the way they spread, such as Recursive Neural Network (RvNN) and so on. However, these deep learning methods only take into account the patterns of deep propagation but ignore the structures of wide dispersion in rumor detection. Actually, propagation and dispersion are two crucial characteristics of rumors.
In this paper, we propose a novel bi-directional graph model, named {\em Bi-Directional Graph Convolutional Networks} (Bi-GCN), to explore both characteristics by operating on both top-down and bottom-up propagation of rumors. It leverages a GCN with a top-down directed graph of rumor spreading to learn the patterns of rumor propagation; and a GCN with an opposite directed graph of rumor diffusion to capture the structures of rumor dispersion. Moreover, the information from source post is involved in each layer of GCN to enhance the influences from the roots of rumors. Encouraging empirical results on several benchmarks confirm the superiority of the proposed method over the state-of-the-art approaches. 
\end{abstract}

\section{Introduction}
With the rapid development of the Internet, social media has become a convenient online platform for users to obtain information, 
express opinions and communicate with each other. As more and more people are keen to participate in discussions about hot 
topics and exchange their opinions on social media, many rumors appear. Due to a large number of users and easy access to social media, rumors can spread widely and quickly on social media, bringing huge harm to society and causing a 
lot of economic losses. 
Therefore, regarding to the potential panic and threat caused by 
rumors, it is urgent to come up with a method to identify rumors on social media efficiently and as early as possible.

Conventional detection methods mainly adopt handcrafted features such as user 
characteristics, text contents and propagation patterns to train supervised classifiers, e.g., Decision Tree \cite{castillo2011information}, 
Random Forest \cite{kwon2013prominent}, Support Vector Machine (SVM) \cite{yang2012automatic}. Some studies apply more 
effective features, such as user comments \cite{giudice2010crowdsourcing}, temporal-structural features \cite{wu2015false}, 
and the emotional attitude of posts \cite{liu2015real}. 
However, those methods mainly rely on feature engineering, which is very time-consuming and labor-intensive. Moreover, those handcrafted features are usually lack of high-level representations extracted from the propagation and the dispersion of rumors.

Recent studies have exploited deep learning methods that mine high-level representations from propagation path/trees or networks to identify rumors. 
Many deep learning models such as Long Short Term Memory (LSTM), Gated Recurrent Unit (GRU), and Recursive Neural Networks (RvNN) 
\cite{ma2016detecting,ma2018rumor} are employed since they are capable to learn sequential features from rumor propagation along time. However, these 
approaches have a significant limitation on efficiency since temporal-structural features only pay attention to the sequential propagation of rumors 
but neglect the influences of rumor dispersion. The structures of rumor dispersion also indicate some spreading behaviors of rumors. Thus, some studies 
have tried to involve the information from the structures of rumor dispersion by invoking Convolutional Neural Network (CNN) based 
methods \cite{yu2017convolutional,yu2019attention}. CNN-based methods can obtain the correlation features within local neighbors but cannot handle the 
global structural relationships in graphs or trees \cite{bruna2014spectral}. Therefore, the global structural features of rumor dispersion are ignored 
in these approaches. Actually, CNN is not designed to learn high-level representations from structured data but Graph Convolutional Network (GCN) is \cite{kipf2017semi}.

\begin{figure}[htbp]
\centering
\subfigure[UD-GCN]{
\label{fig:ud-gcn}
\begin{minipage}[t]{0.3\linewidth}
\centering
\includegraphics[width=1.1\columnwidth]{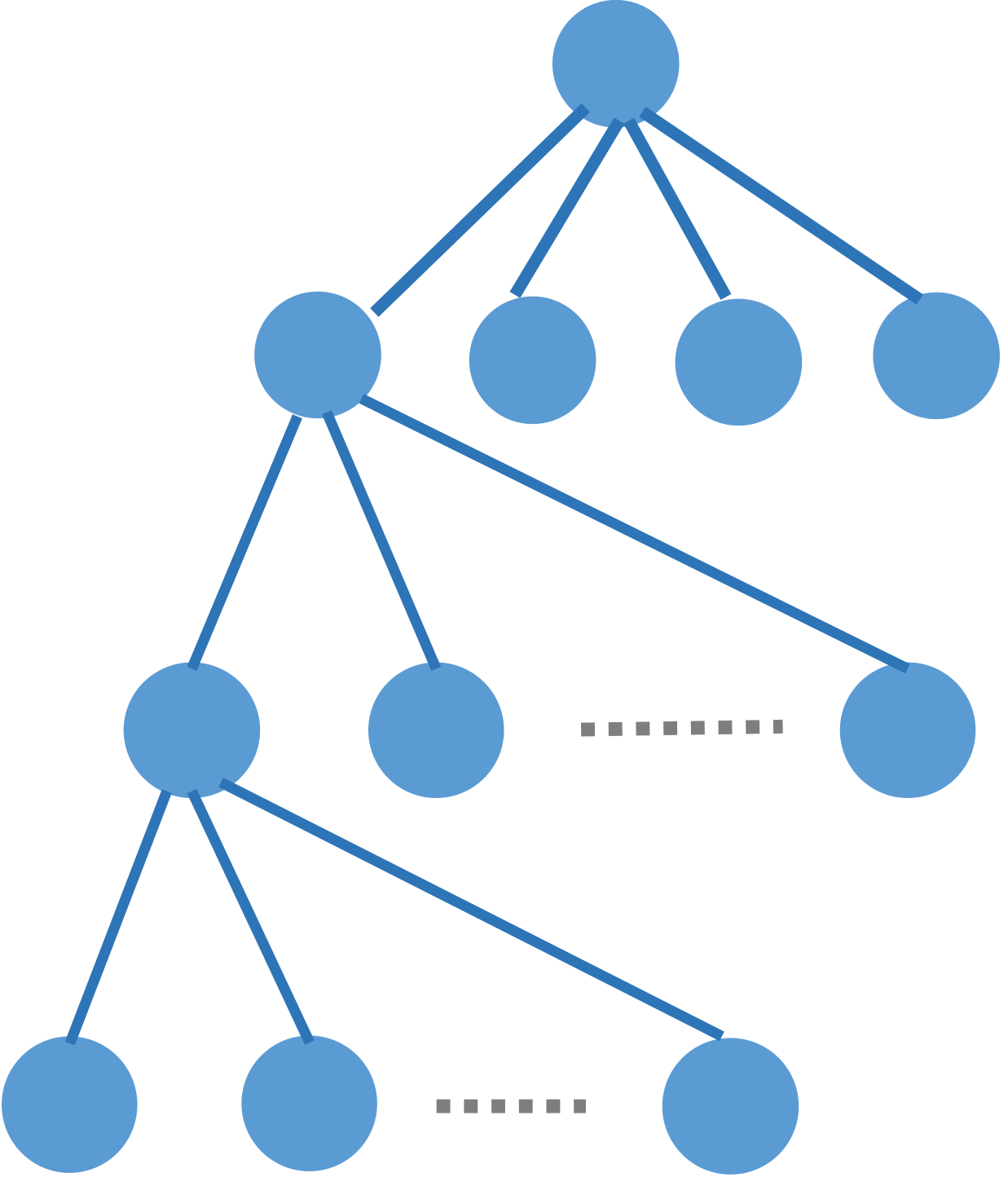}
\end{minipage}
}
\subfigure[TD-GCN]{
\label{fig:td-gcn}
\begin{minipage}[t]{0.3\linewidth}
\centering
\includegraphics[width=1.1\columnwidth]{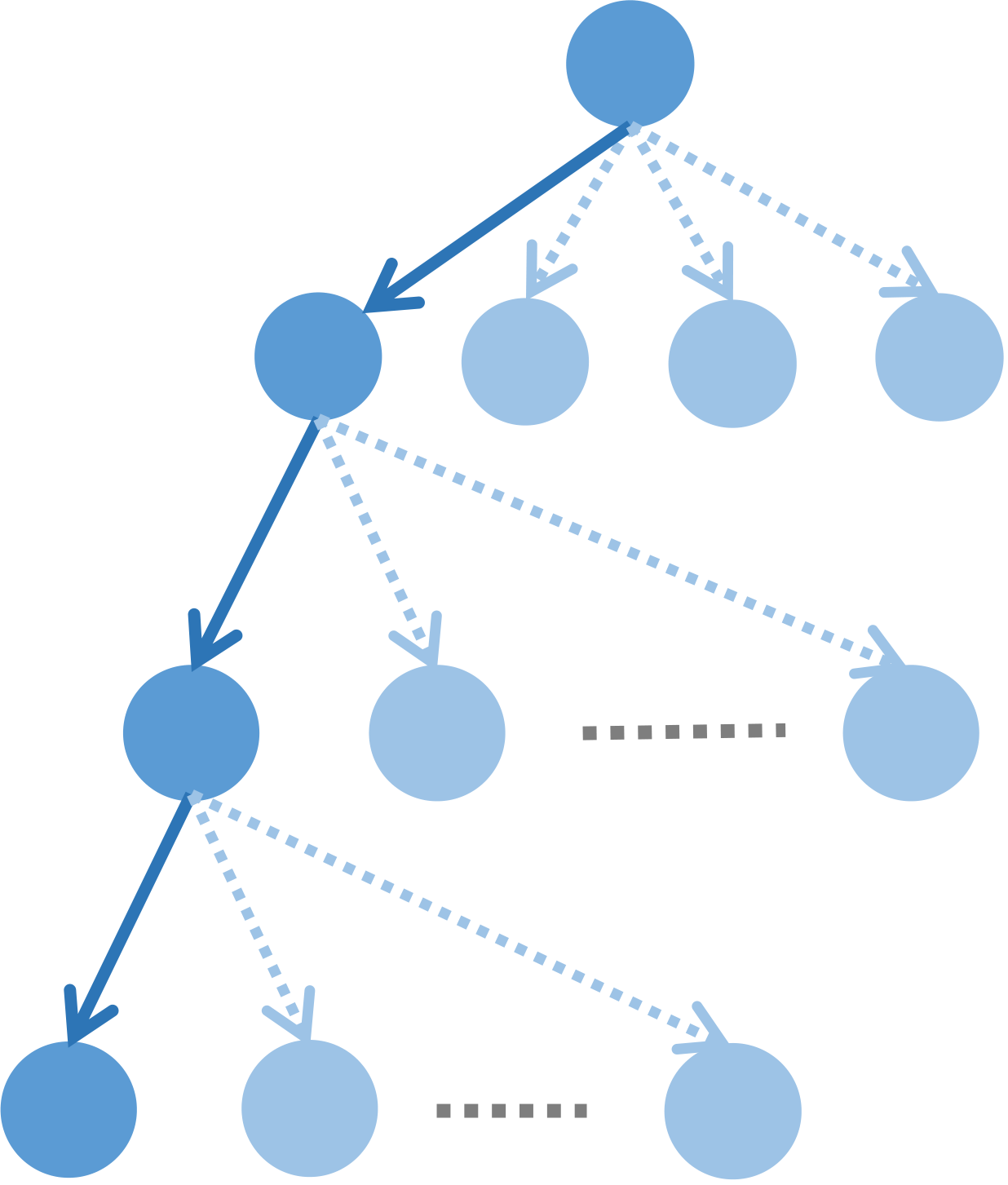}
\end{minipage}
}
\subfigure[BU-GCN]{
\label{fig:bu-gcn}
\begin{minipage}[t]{0.3\linewidth}
\centering
\includegraphics[width=1.1\columnwidth]{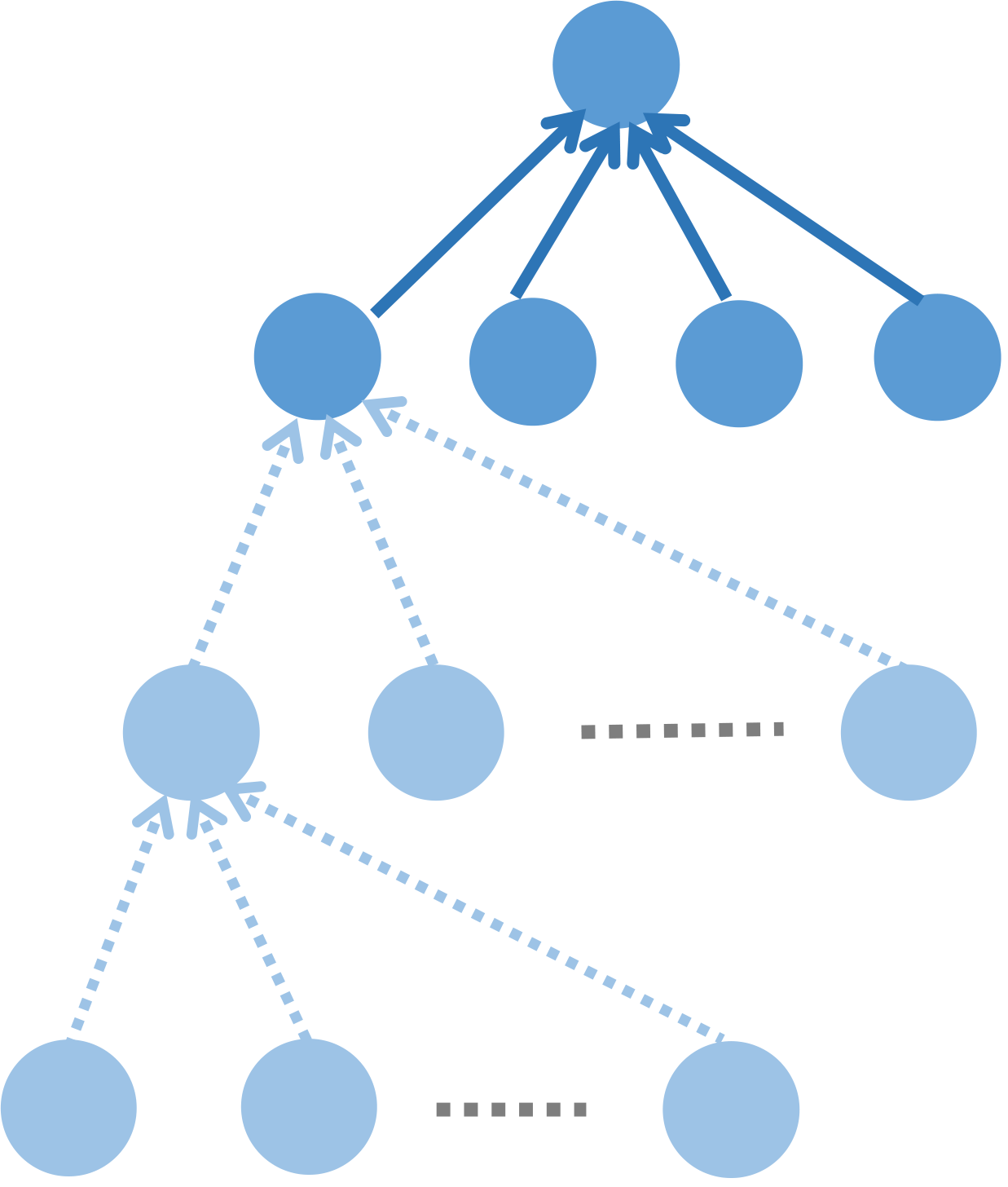}
\end{minipage}
}
\caption{(a) the undirected graph with only node relationships; (b) the deep propagation along a relationship chain from top to down; (c) the aggregation of the wide dispersion within a community to an upper node.}
\label{fig:tree}
\end{figure}

So can we simply apply GCN to rumor detection since it has successfully made progress in various fields, such as social networks \cite{hamilton2017inductive}, 
physical systems \cite{battaglia2016interaction}, and chemical drug discovery \cite{defferrard2016convolutional}? The answer is no. As shown in 
Figure~\ref{fig:ud-gcn}, GCN, or called undirected GCN (UD-GCN), only aggregates information relied on the relationships among relevant posts but loses the 
sequential orders of follows. Although UD-GCN has the ability to handle the global structural features of rumor dispersion, it does not consider the direction 
of the rumor propagation, which however has been shown to be an important clue for rumor detection \cite{wu2015false}.
Specifically, deep propagation along a relationship chain \cite{han2014energy} and wide dispersion across a social community \cite{thomas2007lies} are two major characteristics of rumors, which is eager for a method to serve both.

To deal with both propagation and dispersion of rumors, in this paper, we propose a novel {\em Bi-directional GCN} (Bi-GCN), which operates on both top-down and bottom-up propagation of rumors. The proposed method obtains the features of propagation and dispersion via two parts, the Top-Down graph convolutional Networks (TD-GCN) and Bottom-Up graph convolutional Networks (BU-GCN), respectively. As shown in Figure~\ref{fig:td-gcn} and~\ref{fig:bu-gcn}, TD-GCN forwards information from the parent node of a node in a rumor tree to formulate rumor propagation while BU-GCN aggregates information from the children nodes of a node in a rumor tree to represent rumor dispersion. Then, the representations of propagation and dispersion pooled from the embedding of TD-GCN and BU-GCN are merged together through full connections to make the final results. Meanwhile, we concatenate the features of the roots in rumor trees with the hidden features at each GCN layer to enhance the influences from the roots of rumors. Moreover, we employ DropEdge \cite{rong2019the} in the training phase to avoid over-fitting issues of our model. The main contributions of this work are as follows:
\begin{itemize}
\item We leverage Graph Convolutional Networks to detect rumors. To the best of our knowledge, this is the first study of employing GCN in rumor detection of social media. 
\item We propose the Bi-GCN model that not only considers the causal features of rumor propagation along relationship chains from top to down but also obtains the structural features from rumor dispersion within communities through the bottom-up gathering.
\item We concatenate the features of the source post with other posts at each graph convolutional layer to make a comprehensive use of the information from the root feature and achieve excellent performance in rumor detection.
\end{itemize}

Experimental results on three real-world datasets show that our Bi-GCN method outperforms several state-of-the-art approaches; and for the task of early detection of rumors, which is quite crucial to identify rumors in real time and prevent them from spreading, Bi-GCN also achieves much higher effectiveness.

\section{Related Work}
In recent years, automatic rumor detection on social media has attracted a lot of attention. Most previous work for rumor detection mainly focuses
on extracting rumor features from the text contents, user profiles and propagation structures to learn a classifier from labeled data \cite{castillo2011information,yang2012automatic,kwon2013prominent,liu2015real,zhao2015enquiring}. Ma et al. \cite{ma2015detect} classified the rumor by using the time-series to model the variation of handcrafted social context features. Wu et al. \cite{wu2015false} proposed a graph kernel-based hybrid SVM classifier by combining the RBF kernel with a random-walk-based graph kernel. Ma et al. \cite{ma2017detect} constructed a propagation tree kernel to detect rumors by evaluating the similarities between their propagation tree structures. These methods not only were ineffective but also heavily relied on handcrafted feature engineering to extract informative feature sets.

In order to automatically learn high-level features, a number of recent methods were proposed to detect rumor based on deep learning models. Ma et al. utilized 
Recurrent Neural Networks (RNN) to capture the hidden representation from temporal content features \cite{ma2016detecting}. Chen et al. \cite{chen2018call} 
improved this approach by combining attention mechanisms with RNN to focus on text features with different attentions. 
Yu et al. \cite{yu2017convolutional} proposed a method based on Convolutional Neural Network (CNN) to learn key features scattered 
among an input sequence and shape high-level interactions among significant features. Liu et al. \cite{liu2018early} incorporated both RNN and CNN to get the 
user features based on time series. Recently, Ma et al. \cite{ma2019detect} employed the adversarial learning method to improve the performance of rumor 
classifier, where the discriminator is used as a classifier and the corresponding generator improves the discriminator by generating conflicting noises. 
In addition, Ma et al. built a tree-structured Recursive Neural Networks (RvNN) to catch the hidden representation from 
both propagation structures and text contents \cite{ma2018rumor}. However, these methods are too inefficient to learn the features of the propagation structure, and they also ignore the global structural features of rumor dispersion.

Compared to the deep-learning models mentioned above, GCN is able to capture global structural features from graphs or trees better. Inspired by the success of CNN in the field of computer vision, GCN has demonstrated state-of-the-art performances in various tasks with graph data \cite{battaglia2016interaction,defferrard2016convolutional,hamilton2017inductive}. Scarselli et al. \cite{scarselli2008graph} firstly introduced GCN as a special massage-passing model for either undirected graphs or directed graphs. 
Later on, Bruna et al. \cite{bruna2014spectral} theoretically analyzed graph convolutional methods for undirected graphs based on the spectral graph theory. Subsequently, Defferrard et al. \cite{defferrard2016convolutional} developed a 
method named the Chebyshev Spectral CNN (ChebNet) and used the Chebyshev polynomials as the filter. 
After this work, Kipf et al. \cite{kipf2017semi} presented a first-order approximation of ChebNet (1stChebNet), where the information of each node 
is aggregated from the node itself and its neighboring nodes. Our rumor detection model is inspired by the GCN.

\section{Preliminaries}
We introduce some fundamental concepts that are necessary for our method. First the notation used in this paper is as follows.

\subsection{Notation}
Let $C=\{c_1, c_2, ... , c_m\}$ be the rumor detection dataset, where $c_i$ is the $i$-th event and $m$ is the number of events.
$c_i =\{r_i, w^i_1, w^i_2, ... , w^i_{n_i-1}, G_i\}$, where $n_i$ refers to the number of posts in $c_i$, $r_i$ is the
source post, each $w^i_j$ represents the $j$-th relevant responsive post, and $G_i$ refers to the propagation structure.
Specifically, $G_i$ is defined as a graph $\left\langle V_i, E_i\right\rangle$ with $r_i$ being the root node \cite{wu2015false,ma2017detect}, where $V_i=\{r_i,w_1^i,\ldots,w_{n_i-1}^i\}$, and $E_i=\{e^i_{st}|s,t=0,\ldots,n_i-1\}$ that represents the set of edge from responded posts to the retweeted posts or responsive posts, as shown in Figure \ref{fig:td-gcn}. 
For example, if $w^i_2$ has a response to $w^i_1$, there will be an directed edge   $w^i_1 \to w^i_2$, i.e., $e^i_{12}$. If $w^i_1$ has a response to $r_i$, there will be an directed edge   $r_i \to w^i_1$, i.e., $e^i_{01}$. Denote $\textbf{A}_i\in \{0,1\}^{n_i\times n_i}$ as an adjacency matrix where
\begin{equation*}
a^i_{ts}=
\begin{cases}
1, & \text{if}\;e^i_{st} \in E_i \\
0, & \text{otherwise}
\end{cases}.
\end{equation*}
Denote $\textbf{X}_i=[\textbf{x}_0^{i\top}, \textbf{x}_1^{i\top}, ... , \textbf{x}_{n_i-1}^{i\top}]^\top$ as a feature matrix extracted from the posts in $c_i$, where $\textbf{x}^i_0$ represents the feature vector of $r_i$ and each other row feature $\textbf{x}^i_j$ represents the feature vector of $w^i_j$.

Moreover, each event $c_i$ is associated with a ground-truth label $y_i \in \{F, T\}$  (i.e., False Rumor or True Rumor). In some cases, the label $y_i$ is one of the four finer-grained classes $\{N, F, T, U\}$ (i.e., Non-rumor, False Rumor, True Rumor, and Unverified Rumor)~\cite{ma2017detect,zubiaga2018detection}. 
Given the dataset, the goal of rumor detection is to learn a classifier 
$$
f: C \to Y,
$$
where $C$ and $Y$ are the sets of events and labels respectively, to predict the label of an event based on text contents, user information and propagation structure constructed by the related posts from that event.

\subsection{Graph Convolutional Networks}
Recently, there is an increasing interest in generalizing convolutions to the graph domain. 
Among all the existing works, GCN is one of the most effective convolution models, whose convolution operation is considered as a general "message-passing" architecture as follows:
\begin{equation}
    \textbf{H}_{k}=M(\textbf{A},\textbf{H}_{k-1};\bm{W}_{k-1}),
\end{equation}
where $\textbf{H}_{k}\in\mathbb{R}^{n\times v_k}$ is the hidden feature matrix computed by the $k-th$ Graph Conventional Layer (GCL) and $M$ is the message
propagation function, which depends on the adjacency matrix $\textbf{A}$, the hidden feature matrix $\textbf{H}_{k-1}$ and the trainable parameters $\bm{W}_{k-1}$. 

There are many kinds of message propagation functions $M$ for GCN~\cite{bruna2014spectral,defferrard2016convolutional}. 
Among them, the message propagation function defined in the first-order approximation of ChebNet (1stChebNet)~\cite{kipf2017semi} is
 as follows:
\begin{equation}
\label{equ:gcn}
    \textbf{H}_{k}=M(\textbf{A},\textbf{H}_{k-1};\bm{W}_{k-1})=\sigma(\hat{\textbf{A}}\textbf{H}_{k-1}\bm{W}_{k-1}).
\end{equation}
In the above equation $\hat{\textbf{A}}=\tilde{\textbf{D}}^{-\frac{1}{2}}\tilde{\textbf{A}}\tilde{\textbf{D}}^{-\frac{1}{2}}$ is the normalized adjacency 
matrix, where $\tilde{\textbf{A}}$=$\textbf{A}+\textbf{I}_N$ (i.e., adding self-connection), $\tilde{\textbf{D}}_{ii}$=$\Sigma_j\tilde{\textbf{A}}_{ij}$ that 
represents the degree of the $i-{th}$ node; $\bm{W}_{k-1}\in\mathbb{R}^{v_{k-1}\times v_{k}}$; and $\sigma(\cdot)$ is an activation function, e.g., the ReLU function.

\begin{figure*}[ht]
    \centering
    \includegraphics[width=2\columnwidth]{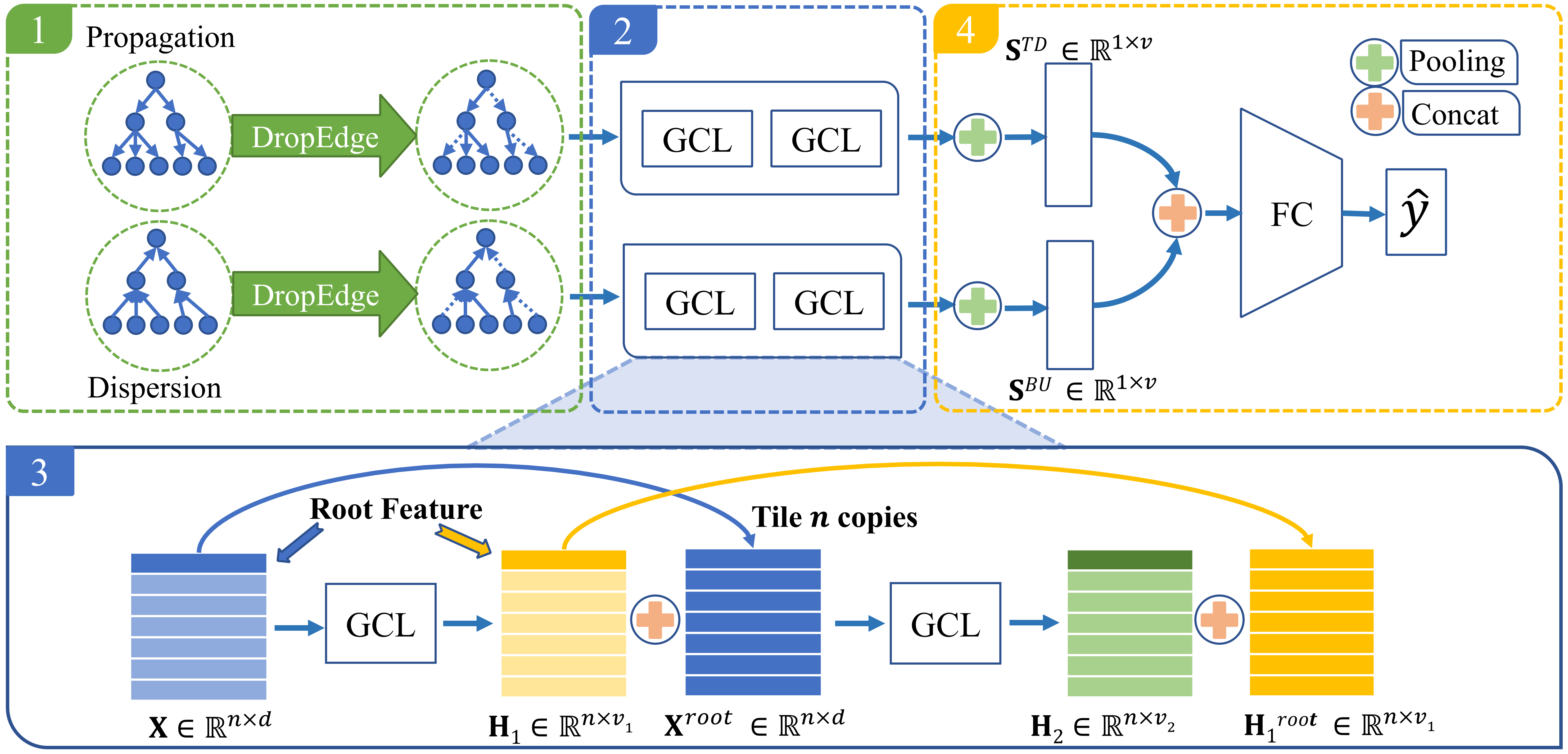}
    \caption{Our Bi-GCN rumor detection model. $\textbf{X}$ denotes the original feature matrix input to the Bi-GCN model, 
    and $\textbf{H}_k$ is the hidden features matrix generated from the $k-{th}$ GCL. 
    $\textbf{X}^{root}$ and $\textbf{H}_1^{root}$ represents the matrix extended by the features of source post.} 
    \label{fig:flowchart}
\end{figure*}

\subsection{DropEdge}
DropEdge is a novel method to reduce over-fitting for GCN-based models~\cite{rong2019the}. 
In each training epoch, it randomly drops out edges from the input graphs to generate different deformed copies with certain rate.
As a result, this method augments the randomness and the diversity of input data, just like rotating or flapping images at random.
Formally, suppose the total number of edges in the graph $\textbf{A}$ is $N_e$ and the dropping rate is $p$, then the adjacency matrix after DropEdge, $\bm{A}'$, is computed as below:
\begin{equation}
\label{equ:dropedge}
    \textbf{A}'=\textbf{A}-\textbf{A}_{drop}
\end{equation}
where $\textbf{A}_{drop}$ is the matrix constructed using $N_e\times p$ edges randomly sampled from the original edge set.

\section{Bi-GCN Rumor Detection Model}
In this section, we propose an effective GCN-based method for rumor detection based on the rumor propagation and the rumor dispersion, named as 
{\em Bi-directional Graph Convolutional Networks} (Bi-GCN).
The core idea of Bi-GCN is to learn suitable high-level representations from both rumor propagation and rumor dispersion.
In our Bi-GCN model, two-layer 1stChebNet are adopted as the fundamental GCN components. As shown in Figure \ref{fig:flowchart}, we elaborate the rumor detection process using Bi-GCN in 4 steps. 

We first discuss how to apply the Bi-GCN model to one event, i.e., $c_i \to y_i$ for the $i$-th event. The other events are calculated in the same manner. To better present our method, we omit the subscript $_i$ in the following content.

\subsection{1 Construct Propagation and Dispersion Graphs}
Based on the retweet and response relationships, we construct the propagate structure $\left\langle V, E\right\rangle$ for a rumor event $c_i$. Then, let $\textbf{A}\in \mathbb{R}^{n_i\times n_i}$ and $\textbf{X}$ be its corresponding adjacency matrix and feature matrix of $c_i$ based on the spreading tree of rumors, respectively. 
$\textbf{A}$ only contains the edges from the upper nodes to the lower nodes as illustrated in Figure~\ref{fig:td-gcn}. 
At each training epoch, $p$ percentage of edges are dropped via Eq. (\ref{equ:dropedge}) to form $\textbf{A}'$, which avoid penitential over-fitting issues \cite{rong2019the}.
Based on $\textbf{A}'$ and $\textbf{X}$, we can build our Bi-GCN model. 
Our Bi-GCN consists of two components: a Top-Down Graph Convolutional Network (TD-GCN) and a Bottom-Up Graph Convolutional Network (BU-GCN). The adjacency matrices of two components are different.
For TD-GCN, the adjacency matrix is represented as $\textbf{A}^{TD}=\textbf{A}'$. Meanwhile, for BU-GCN, the adjacency matrix is $\textbf{A}^{BU}=\textbf{A}'^\top$. 
TD-GCN and BU-GCN adopt the same feature matrix \textbf{X}.

\subsection{2 Calculate the High-level Node Representations} 
After the DropEdge operation, the top-down propagation features and the bottom-up propagation features are obtained by TD-GCN and BU-GCN, respectively.

By substituting $\textbf{A}^{TD}$ and $\textbf{X}$ to Eq. (\ref{equ:gcn}) over two layers, we write the equations for TD-GCN as below:
\begin{equation}\label{equ:hidden1}
    \textbf{H}_1^{TD}=\sigma\left(\hat{\textbf{A}}^{TD}\textbf{X}\bm{W}^{TD}_{0}\right),
\end{equation}
\begin{equation}\label{equ:hidden2}
    \textbf{H}_2^{TD}=\sigma\left(\hat{\textbf{A}}^{TD}\textbf{H}_1^{TD}\bm{W}^{TD}_{1}\right),
\end{equation} 
where $\textbf{H}_1^{TD}\in\mathbb{R}^{n\times v_1}$ and $\textbf{H}_2^{TD}\in\mathbb{R}^{n\times v_2}$ 
represent the hidden features of two layer TD-GCN. $\bm{W}^{TD}_{0}\in\mathbb{R}^{d\times v_1}$ and $\bm{W}^{TD}_{1}\in\mathbb{R}^{v_1\times v_2}$ are the 
filter parameter matrices of TD-GCN. Here we adopt ReLU function as the activation function, $\sigma(\cdot)$. Dropout \cite{srivastava2014dropout} is 
applied on GCN Layers (GCLs) to avoid over-fitting. Similar to Eqs. (\ref{equ:hidden1}) and (\ref{equ:hidden2}), we calculate the bottom-up hidden features $\textbf{H}_1^{BU}$ and $\textbf{H}_2^{BU}$ for BU-GCN in the same manner as Eq. (\ref{equ:hidden1}) and Eq. (\ref{equ:hidden2}).

\subsection{3 Root Feature Enhancement}
As we know, the source post of a rumor event always has abundant information to make a wide impact. It is necessary to better make use of the information from the source post, and learn more accurate node representations from the relationship between nodes and the source post.

Consequently, besides the hidden features from TD-GCN and BU-GCN, we propose an operation of root feature enhancement to improve the performance of rumor detection as shown in Figure \ref{fig:flowchart}. 
Specifically, for TD-GCN at the $k$-th GCL, we concatenate the hidden feature vectors of every 
nodes with the hidden feature vector of the root node from the $(k-1)$-th GCL to construct a new feature matrix as
\begin{equation}
 \tilde{\textbf{H}}^{TD}_{k}=\text{concat}(\textbf{H}^{TD}_{k},(\textbf{H}^{TD}_{k-1})^{root})   
\end{equation}
with $\textbf{H}^{TD}_{0}=\textbf{X}$.   
Therefore, we express TD-GCN with the root feature enhancement by replacing $\textbf{H}_{1}^{TD}$ in Eq. (\ref{equ:hidden2}) with $\tilde{\textbf{H}}^{TD}_{1}=\text{concat}(\textbf{H}_{1}^{TD},\textbf{X}^{root})$, and then get $\tilde{\textbf{H}}^{TD}_{2}$ as follows:
\begin{equation}\label{equ:root_hidden1}
    \textbf{H}_2^{TD}=\sigma\left(\hat{\textbf{A}}^{TD}\tilde{\textbf{H}}_{1}^{TD}\bm{W}^{TD}_{1}\right),
\end{equation}
\begin{equation}\label{equ:root_hidden2}
    \tilde{\textbf{H}}_{2}^{TD}=\text{concat}(\textbf{H}_{2}^{TD},(\textbf{H}^{TD}_{1})^{root}).
\end{equation}
Similarly, the hidden feature metrics of BU-GCN with root feature enhancement, $\tilde{\textbf{H}}_{1}^{BU}$ and $\tilde{\textbf{H}}_{2}^{BU}$, are obtained in the same manner as Eq. (\ref{equ:root_hidden1}) and Eq. (\ref{equ:root_hidden2}).

\subsection{4 Representations of Propagation and Dispersion for Rumor Classification} 
The representations of propagation and dispersion are the aggregations from the node representations of TD-GCN and BU-GCN, respectively. Here we employ mean-pooling operators to aggregate information from these two sets of the node representations. It is formulated as 

\begin{equation}
    \textbf{S}^{TD} =\text{MEAN}(\tilde{\textbf{H}}_{2}^{TD}),
\end{equation}
\begin{equation}
    \textbf{S}^{BU}=\text{MEAN}(\tilde{\textbf{H}}_{2}^{BU}).
\end{equation}
Then, we concatenate the representations of propagation and the representation of dispersion to merge the information as 
\begin{equation}
    \textbf{S}=\text{concat}(\textbf{S}^{TD}, \textbf{S}^{BU}).
\end{equation}
Finally, the label of the event $\hat{\textbf{y}}$ is calculated via several full connection layers and a softmax layer:
\begin{equation}
    \hat{\textbf{y}}=Softmax(FC(\textbf{S})).
\end{equation}
where $\hat{\textbf{y}}\in\mathbb{R}^{1\times C}$ is a vector of probabilities for all the classes used to predict the label of the event.

We train all the parameters in the Bi-GCN model by minimizing the cross-entropy of the predictions and ground truth distributions, $Y$, 
over all events, $C$. $L_2$ regularizer is applied in the loss function over all the model parameters.

\section{Experiments}
In this section, we first evaluate the empirical performance of our proposed Bi-GCN method in comparison with several baseline models.  
Then, we investigate the effect of each variant of the proposed method.
Finally, we also examine the capability of early rumor detection for both the proposed method and the compared methods.

\subsection{Settings and Datasets}
\subsubsection{Datasets}

We evaluate our proposed method on three real-world datasets: \textit{Weibo} \cite{ma2016detecting}, \textit{Twitter15} \cite{ma2017detect}, and \textit{Twitter16} \cite{ma2017detect}. 
\textit{Weibo} and \textit{Twitter} are the most popular social media sites in China and the U.S., respectively. 
In all the three datasets, nodes refer to users, edges represent retweet or response relationships, and features are the extracted top-5000 words in terms of the 
TF-IDF values as mentioned in the Bi-GCN Rumor Detection Model Section.
The \textit{Weibo} dataset contains two binary labels: False Rumor (F) and True Rumor (T), while \textit{Twitter15} and \textit{Twitter16} datasets contains four labels: Non-rumor (N), False Rumor (F), True Rumor (T), and Unverified Rumor (U). 
The label of each event in \textit{Weibo} is annotated according to Sina community management center, which reports various misinformation \cite{ma2016detecting}. And the label of each event in \textit{Twitter15} and \textit{Twitter16} is annotated according to the veracity tag of the article in rumor debunking websites (e.g., snopes.com, Emergent.info, etc) \cite{ma2017detect}.
The statistics of the three datasets are shown in Table~\ref{table1}.

\begin{table}[t]
    \caption{Statistics of the datasets}\smallskip
    \centering
    \resizebox{0.47\textwidth}{!}{ 
    \begin{tabular}{l|lll}
    \toprule    
    Statistic           & \textit{Weibo}     & \textit{Twitter15} & \textit{Twitter16} \\
    \midrule
    \midrule
    \# of posts            & 3,805,656 & 331,612   & 204,820   \\
    \midrule
    \# of Users            & 2,746,818 & 276,663   & 173,487   \\
    \midrule
    \# of events           & 4664      & 1490      & 818       \\
    \midrule
    \# of True rumors       & 2351      & 374       & 205       \\
    \midrule
    \# of False rumors      & 2313      & 370       & 205       \\
    \midrule
    \# of Unverified rumors & 0         & 374       & 203       \\
    \midrule
    \# of Non-rumors        & 0         & 372       & 205       \\
    \midrule
    Avg. time length / event &  2,460.7 Hours  & 1,337 Hours & 848 Hours \\
    \midrule
    Avg. \# of posts / event & 816  & 223 & 251    \\
    \midrule
    Max \# of posts / event & 59,318  & 1,768 & 2,765    \\
    \midrule
    Min \# of posts / event &  10 & 55 & 81    \\
    \bottomrule
\end{tabular}
}
\label{table1}
\end{table}    

\subsubsection{Experimental Setup}
We compare the proposed method with some state-of-the-art baselines, including:
\begin{itemize}
\item DTC \cite{castillo2011information}: A rumor detection method using a Decision Tree classifier based on various handcrafted 
features to obtain information credibility.
\item SVM-RBF \cite{yang2012automatic}: A SVM-based model with RBF kernel, using handcrafted features based on the overall 
statistics of the posts.
\item SVM-TS \cite{ma2015detect}: A linear SVM classifier that leverages handcrafted features to construct time-series 
model.
\item SVM-TK \cite{ma2017detect}: A SVM classifier with a propagation Tree Kernel on the basis of the propagation structures of rumors.
\item RvNN \cite{ma2018rumor}: A rumor detection approach based on tree-structured recursive neural networks with GRU units that 
learn rumor representations via the propagation structure.
\item PPC\_RNN+CNN \cite{liu2018early}: A rumor detection model combining RNN and CNN, which learns the rumor representations through 
the characteristics of users in the rumor propagation path.
\item Bi-GCN: Our GCN-based rumor detection model utilizing the Bi-directional propagation structure.
\end{itemize}

We implement DTC and SVM-based models with scikit-learn\footnote[1]{https://scikit-learn.org}; PPC\_RNN+CNN with 
Keras\footnote[2]{https://keras.io/}; RvNN and our method with Pytorch\footnote[3]{https://pytorch.org/}.
To make a fair comparison, we randomly split the datasets into five parts, and conduct 5-fold cross-validation to obtain robust results.
For the \textit{Weibo} dataset, we evaluate the Accuracy (Acc.) over the two categories and Precision (Prec.), Recall (Rec.), F1 measure ($F_1$) on each class. 
For the two \textit{Twiter} datasets, we evaluate Acc. over the four categories and $F_1$ on each class. The parameters of Bi-GCN are updated using stochastic gradient descent, 
and we optimize the model by Adam algorithm \cite{kingma2014adam}. The dimension of each node's hidden feature vectors are 64. The dropping rate in DropEdge is 0.2 and the rate of dropout is 0.5.
The training process is iterated upon 200 epochs, and early stopping \cite{yao2007early} is applied when the validation loss stops decreasing by 10 epochs. 
Note that we do not employ SVM-TK on the \textit{Weibo} dataset due to its exponential complexity on large datasets.


\subsection{Overall Performance}
\begin{table}[t]
\caption{Rumor detection results on \textit{Weibo} dataset (F: False Rumor; T: True Rumor)}\smallskip
\centering
\resizebox{0.47\textwidth}{!}{ 
\begin{tabular}{c|c|c|c|c|c}
\toprule    
Method       & Class & Acc. & Prec. & Rec.& $F_1$ \\
\midrule
\midrule 
\multirow{2}*{DTC}          &  F     &   \multirow{2}*{0.831}   &    0.847   &    0.815     &   0.831   \\
                            &  T     &       &   0.815    &   0.824      &   0.819   \\
\midrule                        
\multirow{2}*{SVM-RBF}      &  F     &   \multirow{2}*{0.879}   &   0.777    &    0.656     &   0.708   \\
                            &  T     &      &   0.579    &    0.708     &   0.615   \\
\midrule
\multirow{2}*{SVM-TS}       &  F     &   \multirow{2}*{0.885}   &   0.950    &     0.932    &   0.938   \\
                            &  T     &      &    0.124   &    0.047     &  0.059    \\
\midrule
\multirow{2}*{RvNN}         &  F     &   \multirow{2}*{0.908}    &   0.912    &    0.897     &   0.905   \\
                            &  T     &      &   0.904    &    0.918     &   0.911   \\
\midrule
\multirow{2}*{PPC\_RNN+CNN} &  F     &  \multirow{2}*{0.916}     &   0.884    &    0.957     &   0.919   \\
                            &  T     &      &   0.955    &    0.876     &   0.913   \\
\midrule
\multirow{2}*{Bi-GCN}     &  F     &   \multirow{2}*{\textbf{0.961}}     &  \textbf{0.961}   &    \textbf{0.964}    &   \textbf{0.961}   \\
                            &  T     &      &   \textbf{0.962}    &    \textbf{0.962}   &   \textbf{0.960}   \\
\bottomrule
\end{tabular}
}
\label{table2}
\end{table}    

\begin{table}[t]
\caption{Rumor detection results on \textit{Twitter15} and \textit{Twitter16} datasets (N: Non-Rumor; F: False Rumor; T: True Rumor; U: Unverified 
Rumor)}\smallskip
\centering
\resizebox{0.47\textwidth}{!}{ 
\begin{tabular}{c|c|cccc}
\toprule
\multicolumn{6}{c}{\textit{Twitter15}}\\
\midrule
\multirow{2}*{Method}       & \multirow{2}*{Acc.} & N & F & T & U \\
\cmidrule{3-6}
                            &                    & $F_1$& $F_1$& $F_1$& $F_1$\\
\midrule
DTC          &   0.454    &   0.415   &    0.355   &    0.733     &   0.317   \\
\midrule
SVM-RBF      &   0.318    &   0.225   &    0.082   &    0.455     &   0.218   \\
\midrule
SVM-TS       &   0.544    &   0.796   &   0.472    &    0.404     &   0.483   \\
\midrule
SVM-TK       &   0.750    &   0.804   &   0.698    &     0.765    &   0.733   \\
\midrule
RvNN         &   0.723    &   0.682   &   0.758    &     0.821    &   0.654   \\
\midrule
PPC\_RNN+CNN &   0.477    &   0.359   &   0.507  &    0.300    &   0.640  \\
\midrule
Bi-GCN     &   \textbf{0.886}    &   \textbf{0.891}   &   \textbf{0.860}    &    \textbf{0.930}     &   \textbf{0.864}   \\
\bottomrule
\toprule
\multicolumn{6}{c}{\textit{Twitter16}}\\
\midrule
\multirow{2}*{Method}       & \multirow{2}*{Acc.} & N & F & T & U \\
\cmidrule{3-6}
                            &                     & $F_1$& $F_1$& $F_1$& $F_1$\\
\midrule
DTC          &   0.473    &   0.254   &   0.080    &     0.190    &   0.482   \\
\midrule
SVM-RBF      &   0.553    &  0.670    &   0.085    &     0.117    &   0.361   \\
\midrule
SVM-TS       &    0.574   &   0.755   &   0.420    &   0.571      &   0.526   \\
\midrule
SVM-TK       &    0.732   &  0.740    &    0.709   &    0.836     &   0.686   \\
\midrule                        
RvNN         &    0.737   &   0.662   &   0.743    &     0.835    &   0.708   \\
\midrule
PPC\_RNN+CNN &   0.564   &  0.591   &   0.543   &    0.394     &  0.674   \\
\midrule
Bi-GCN     &   \textbf{0.880}    &   \textbf{0.847}   &  \textbf{ 0.869}    &    \textbf{0.937}     &  \textbf{0.865}    \\
\bottomrule
\end{tabular}
}
\label{table3}
\end{table} 
Table \ref{table2} and Table \ref{table3} show the performance of the proposed method and all the compared methods on the \textit{Weibo} and \textit{Twitter} datasets, respectively.

First, among the baseline algorithms, we observe that the deep learning methods performs significantly better than those using hand-crafted features.
It is not surprising, since the deep learning methods are able to learn high-level representations of rumors to capture valid features. 
This demonstrates the importance and necessity of studying deep learning for rumor detection.

Second, the proposed method outperforms the PPC\_RNN+CNN method in terms of all the performance measures, which indicates the effectiveness of incorporating the dispersion structure for rumor detection. 
Since RNN and CNN cannot process data with the graph structure, PPC\_RNN+CNN ignores important structural features of rumor dispersion. 
This prevents it from obtaining efficient high-level representations of rumors, resulting in worse performance on rumor detection.

Finally, Bi-GCN is significantly superior to the RvNN method. Since RvNN only uses the hidden feature vector of all the leaf nodes 
so that it is heavily impacted by the information of the latest posts.
However, the latest posts are always lack of information such as comments, and just follow the former posts.
Unlike RvNN, the root feature enhancement allows the proposed method to pay more attention to the information of the source posts, 
which helps improve our models much more.

\begin{figure*}[t]
    \centering
    \subfigure[\textit{Weibo} dataset]{
    \label{weibobar}
    \begin{minipage}[t]{0.3\linewidth}
    \centering
    \includegraphics[width=1\columnwidth]{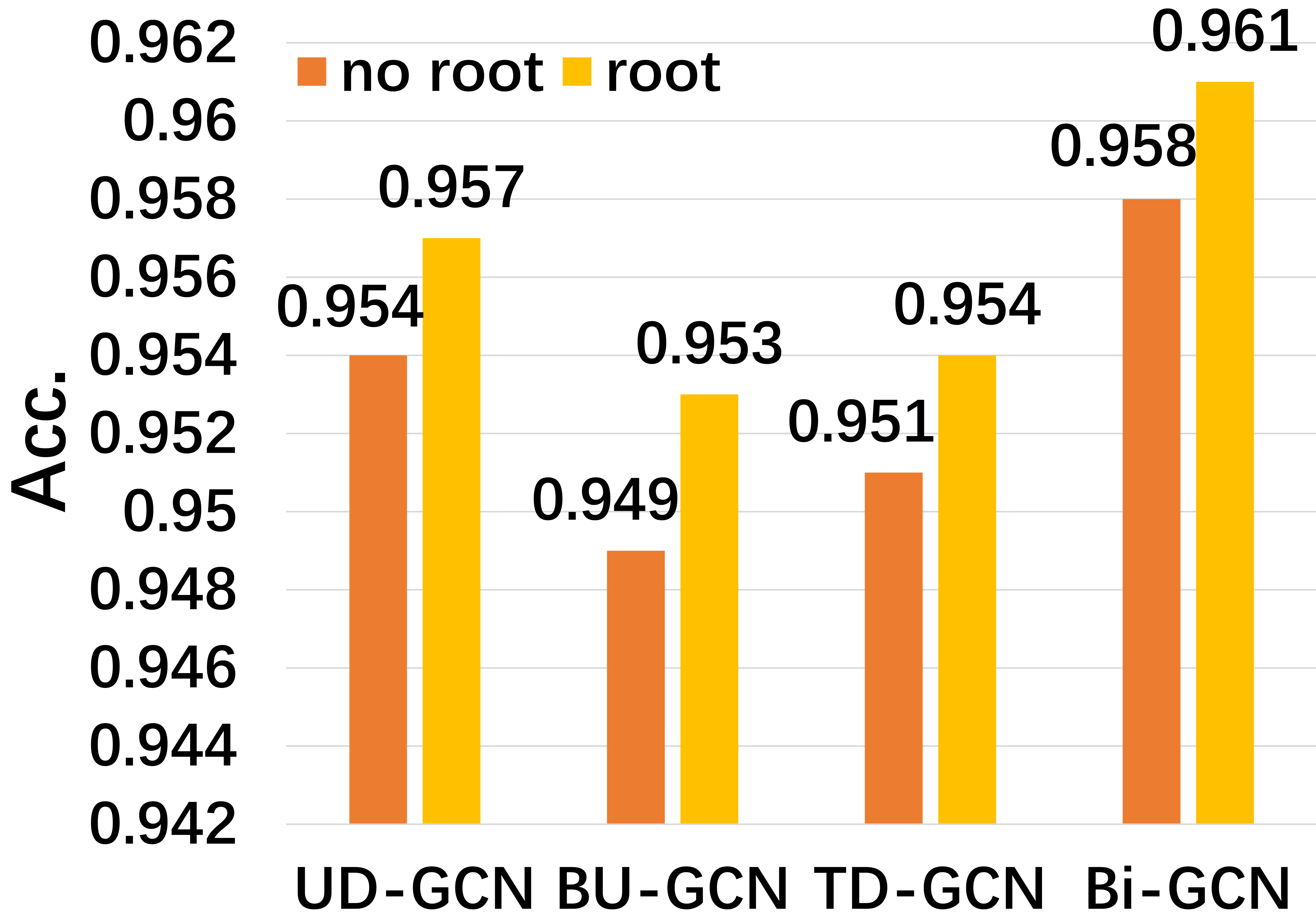}
    \end{minipage}
    }
    \subfigure[\textit{Twitter15} dataset]{
    \label{tw15bar}
    \begin{minipage}[t]{0.3\linewidth}
    \centering
    \includegraphics[width=1\columnwidth]{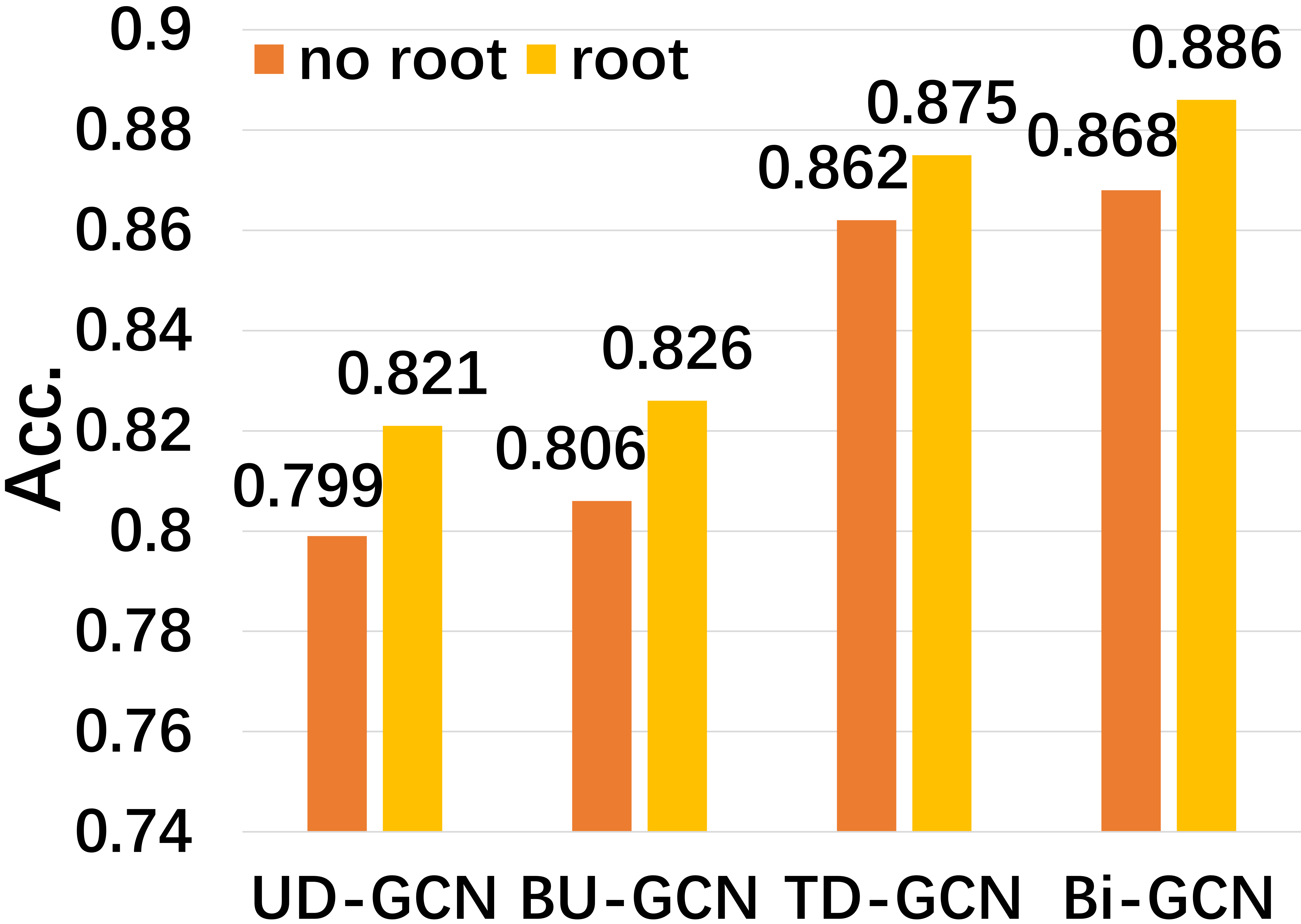}
    \end{minipage}
    }
    \subfigure[\textit{Twitter16} dataset]{
    \label{tw16bar}
    \begin{minipage}[t]{0.3\linewidth}
    \centering
    \includegraphics[width=1\columnwidth]{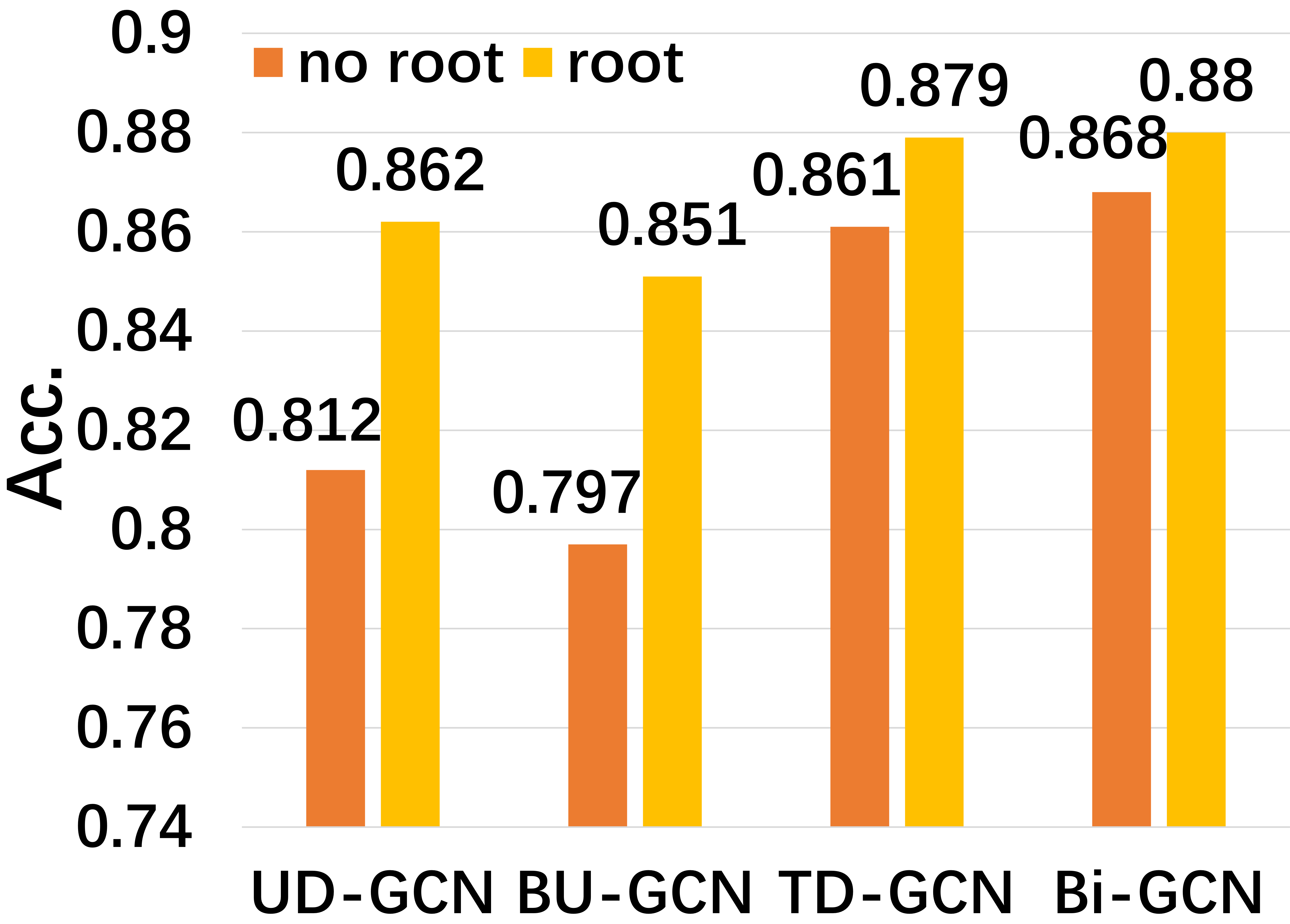}
    \end{minipage}
    }
    \caption{The rumor detection performance of the GCN-based methods on three datasets}
    \label{bar}
    \end{figure*}
\begin{figure*}[t]
\centering
\subfigure[\textit{Weibo} dataset]{
\label{weiboearly}
\begin{minipage}[t]{0.3\linewidth}
\centering
\includegraphics[width=1\columnwidth]{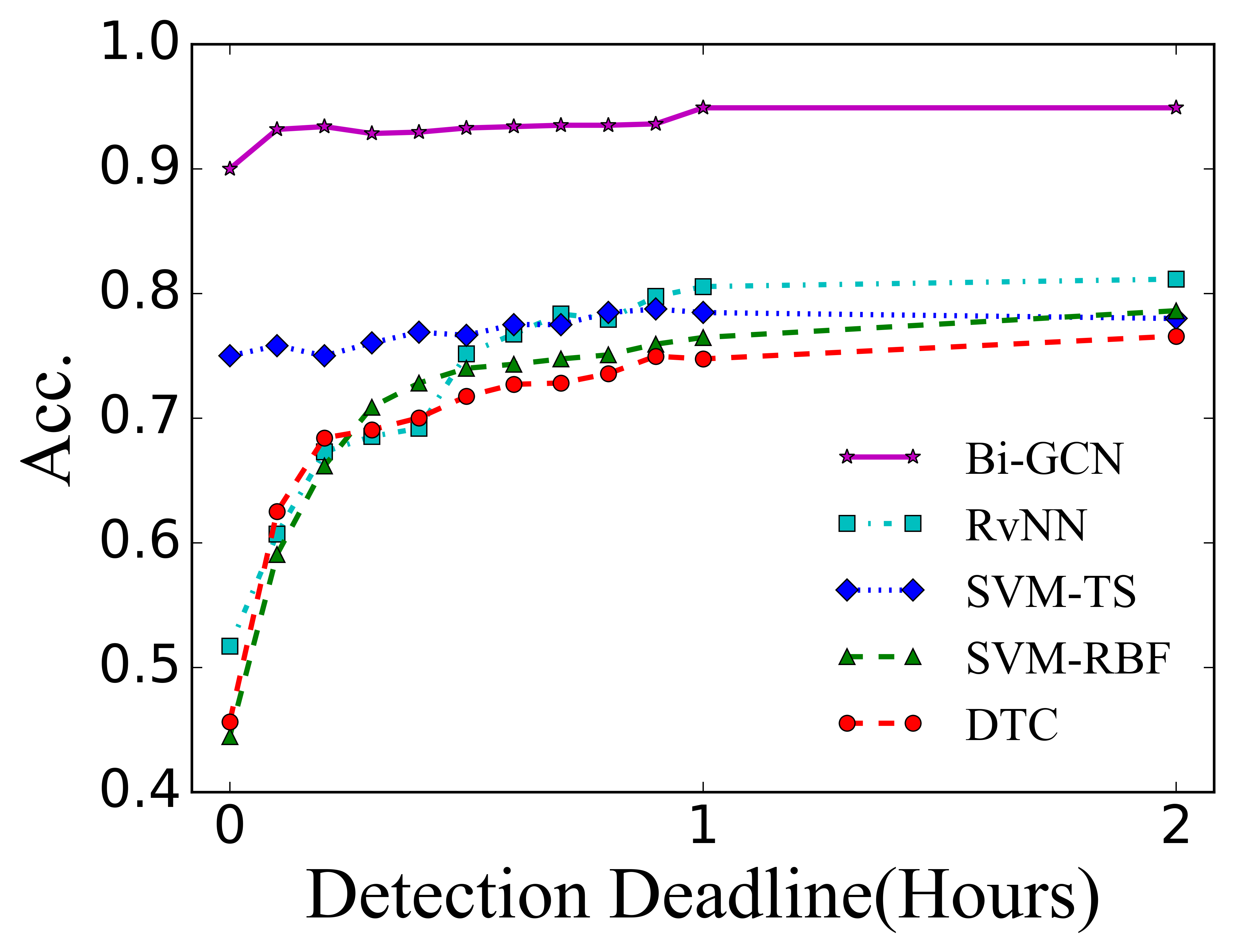}
\end{minipage}
}
\subfigure[\textit{Twitter15} dataset]{
\label{tw15early}
\begin{minipage}[t]{0.3\linewidth}
\centering
\includegraphics[width=1\columnwidth]{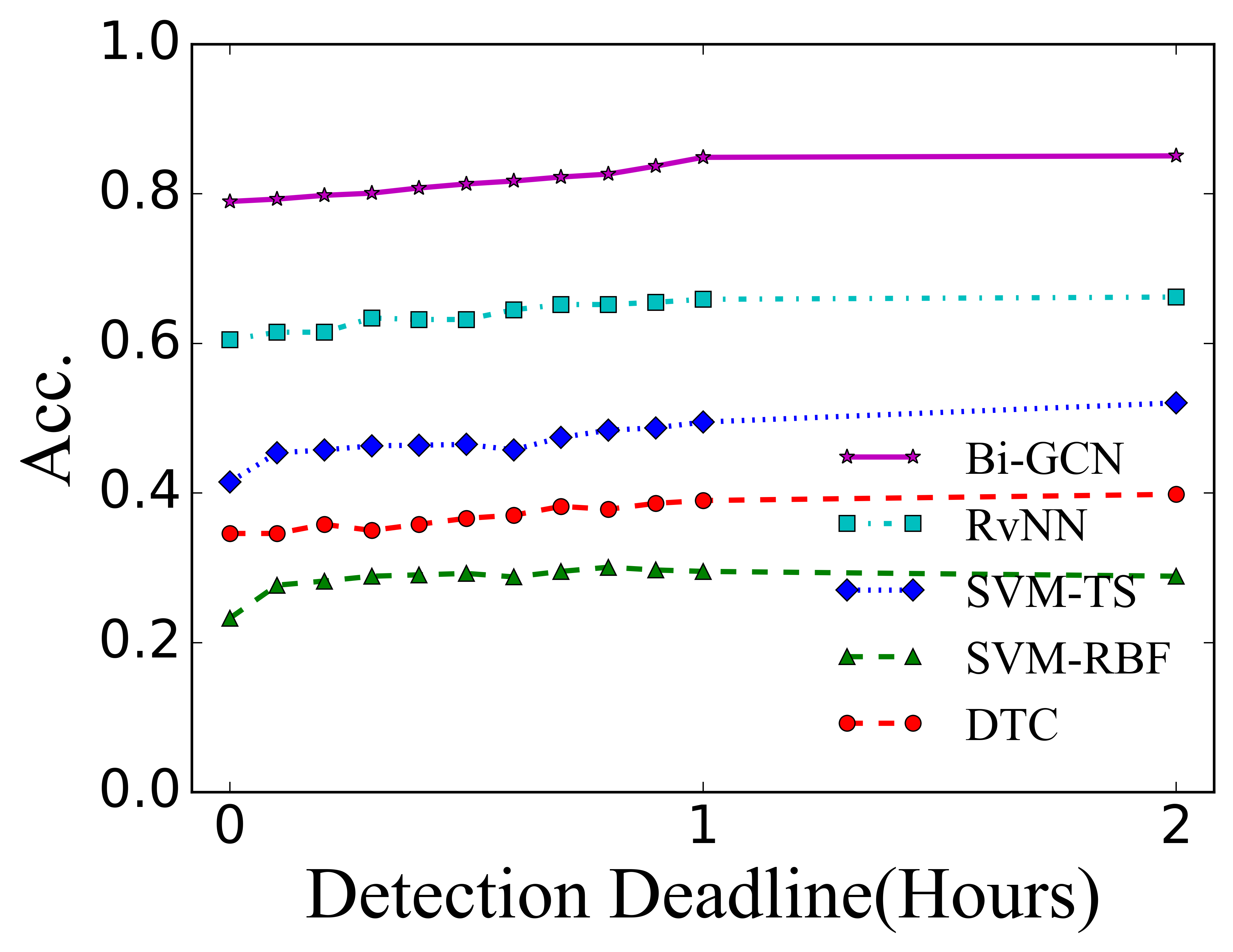}
\end{minipage}
}
\subfigure[\textit{Twitter16} dataset]{
\label{tw16early}
\begin{minipage}[t]{0.3\linewidth}
\centering
\includegraphics[width=1\columnwidth]{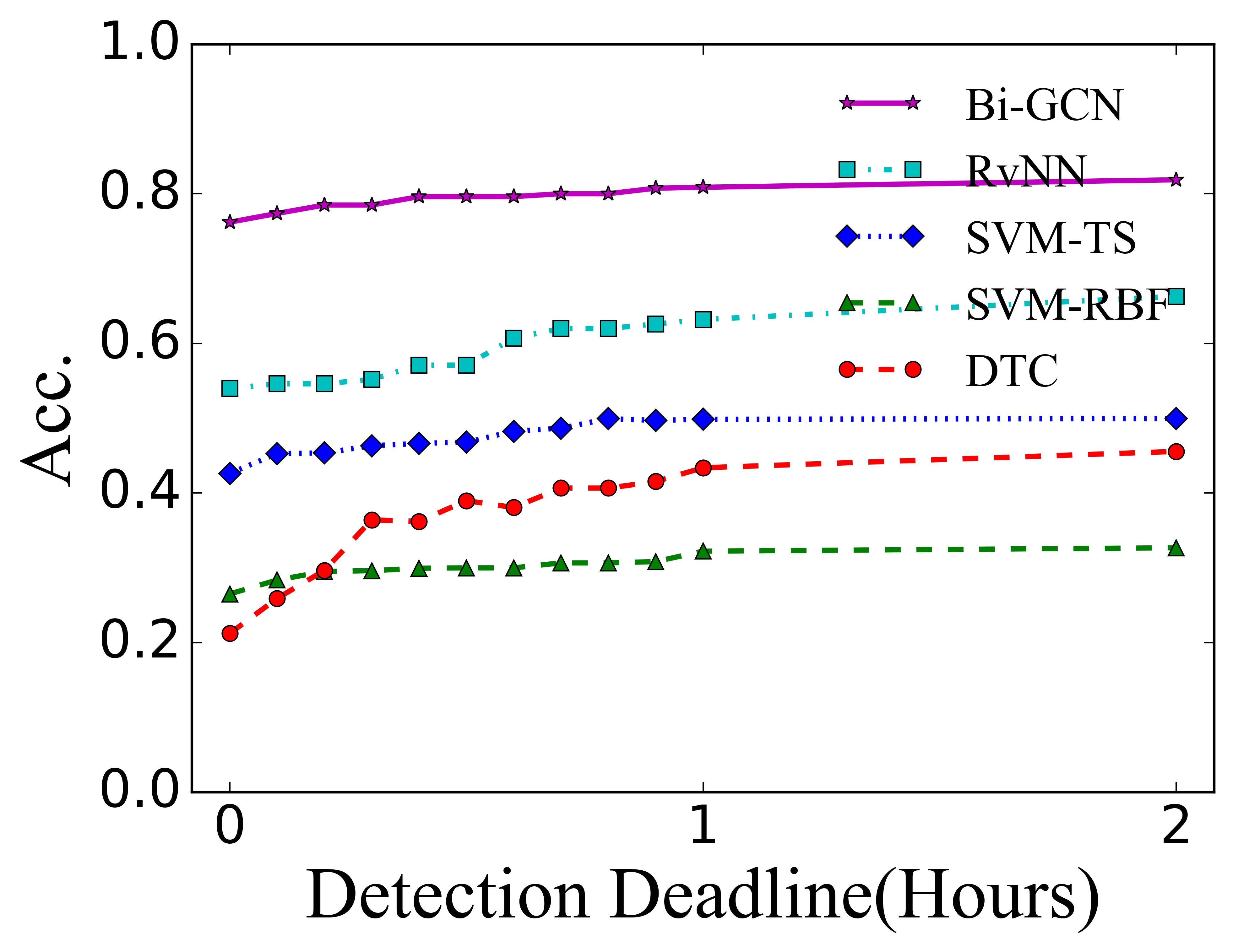}
\end{minipage}
}
\caption{Result of rumor early detection on three datasets}
\label{early}
\end{figure*}
\subsection{Ablation Study}
To analyze the effect of each variant of Bi-GCN, we compare the proposed method with TD-GCN, BU-GCN, UD-GCN and their variants without the root feature 
enhancement. The empirical results are summarized in Figure \ref{bar}. UD-GCN, TD-GCN, and BU-GCN represent our GCN-based rumor detection models utilize 
the UnDirected, Top-Down and Bottom-Up structures, respectively. Meanwhile, "root" refers to the GCN-based model with concatenating root features in the
networks while "no root" represents the GCN-based model without 
concatenating root features in the networks. Some conclusions are drawn from Figure \ref{bar}. First, Bi-GCN, TD-GCN, BU-GCN, and UD-GCN outperforms 
their variants without the root feature enhancement, respectively.
This indicates that the source posts plays an important role in rumor detection.
Second, TD-GCN and BU-GCN can not always achieve better results than UD-GCN, but Bi-GCN is always superior to UD-GCN, TD-GCN and BU-GCN. This implies the importance to simultaneously consider both top-down representations from the ancestor nodes, and bottom-up representations from the children nodes.
Finally, even the worst results in Figures \ref{bar} are better than those of other baseline methods in Table \ref{table2} and \ref{table3} by a large gap, 
which again verifies the effectiveness of graph convolution for rumor detection.

\subsection{Early Rumor Detection}

Early detection aims to detect rumor at the early stage of propagation, which is another important metric to evaluate the quality of the method. 
To construct an early detection task, we set up a series of detection deadlines and only use the posts released before the deadlines to evaluate the accuracy of the proposed method and baseline methods. 
Since it is difficult for the PPC\_RNN+CNN method to process the data of variational lengths, we cannot get the accurate results of PPC\_RNN+CNN at each deadline in this task, so it is not compared in this experiment.


Figure \ref{early} shows the performances of our Bi-GCN method versus RvNN, SVM-TS, SVM-RBF and DTC at various deadlines for the \textit{Weibo} 
and \textit{Twitter} datasets. From the figure, it can be seen that the proposed Bi-GCN method reaches relatively high accuracy at a very early 
period after the source post initial broadcast. Besides, the performance of Bi-GCN is remarkably superior to other models at each deadline, 
which demonstrates that structural features are not only beneficial to long-term rumor detection, but also helpful to the early detection of rumors.

\section{Conclusions}
In this paper, we propose a GCN-based model for rumor detection on social media, called Bi-GCN. 
Its inherent GCN model gives the proposed method the ability of processing graph/tree structures and learning higher-level representations more conducive to 
rumor detection. In addition, we also improve the effectiveness of the model by concatenating the features of the source post after each GCL of GCN. 
Meanwhile, we construct several variants of Bi-GCN to model the propagation patterns, i.e., UD-GCN, TD-GCN and BU-GCN. 
The experimental results on three real-world datasets demonstrate that the GCN-based approaches outperform state-of-the-art baselines in very large margins 
in terms of both accuracy and efficiency.
In particular, the Bi-GCN model achieves the best performance by considering both the causal features of rumor propagation along relationship chains from top to down propagation pattern and the structural features from rumor dispersion within communities through the bottom-up gathering.

\section{Acknowledgments}
The authors would like to thank the support of Tencent AI Lab and Tencent Rhino-Bird Elite Training Program.
This work is supported by the National Natural Science Foundation of Guangdong Province (2018A030313422), National Natural Science Foundation of China (Grant No. 61773229, No. 61972219) and Overseas Cooperation Research Fund of Graduate School at Shenzhen, Tsinghua University (Grant No. HW2018002). 

\small\bibliographystyle{aaai}
\bibliography{Bibliography-File}
\end{document}